\documentclass[aps,preprintnumbers,amsmath,amssymb,nofootinbib]{revtex4}
\usepackage[latin9]{inputenc}
\usepackage{mathtools}
\usepackage{amsmath}
\usepackage{amssymb}
\usepackage{physics}

\usepackage{xcolor}
\usepackage{hyperref}

\date{\today}
\renewcommand{\v}[1]{ \ensuremath{ {\underline{#1}} }}

\begin{document}

\title{Entanglement, partial set of measurements, and diagonality of the
    density matrix in the parton model} 
\author{Haowu Duan}
\affiliation{North Carolina State University, Raleigh, NC 27695, USA}
\author{Candost Akkaya} 
\affiliation{Physics Department, University of Connecticut, 2152 Hillside Road, Storrs, CT 06269, USA}
\author{Alex Kovner}
\affiliation{Physics Department, University of Connecticut, 2152 Hillside Road, Storrs, CT 06269, USA}
\author{Vladimir V. Skokov}
\email{VSkokov@ncsu.edu}
\affiliation{North Carolina State University, Raleigh, NC 27695, USA}
\affiliation{RIKEN-BNL Research Center, Brookhaven National Laboratory, Upton, NY 11973, USA}
\begin{abstract}
We define the ``entropy of ignorance'' which quantifies the entropy associated with  ability to perform only a partial set of measurement on a quantum system. For a parton model the entropy of ignorance is equal to a Boltzmann entropy  of a classical system of partons.
We analyze a calculable model used for describing low x gluons in Color Glass Condensate approach, which has similarities with the parton model of QCD. In this model  we calculate the entropy of ignorance in the particle number basis as well as the entanglement entropy of the observable degrees of freedom. We find that the two are similar at high momenta, but differ by a factor of order unity at low momenta. This holds for the Renyi as well as von Neumann entropies. We conclude that the entanglement  does not seem to play an important role in the context of the parton model.
\end{abstract}  

\maketitle

\section{Introduction} 

In recent years very interesting quantum information theory\footnote{See Ref.~\cite{Witten:2018zva}  for a short introduction.} connections have begun to be explored in the context of high energy and nuclear physics. A set of ideas has been floated which suggests a deep relation between the properties of scattering, such as spectrum of produced particles and entanglement properties of hadronic wave function~\cite{Kutak:2011rb,Peschanski:2012cw,Kovner:2015hga,Peschanski:2016hgk,Berges:2017zws,Hagiwara:2017uaz,Baker:2017wtt,Neill:2018uqw,Liu:2018gae,Feal:2018ptp,Armesto:2019mna,Kovner:2018rbf}. These ideas have found some tentative support in qualitative comparative analysis of data in Ref.~\cite{Tu:2019ouv}. It is thus interesting to elucidate to what extent this way of thinking can be subjected to a more quantitative test.

In this paper, we make a step in this direction. In particular, we ask if the relation suggested in Ref.~\cite{Kharzeev:2017qzs} between the entropy in the parton model and the entropy of entanglement in a proton wave function exists in a computable  model of a hadronic wave function frequently used in the Color Glass Condensate (CGC) calculations (see Refs.~\cite{Iancu:2002xk,Kovner:2005pe,McLerran:2008uj,Gelis:2010nm,Kovchegov:2012mbw} for reviews on CGC).

The authors of  Ref.~\cite{Kharzeev:2017qzs} considered  the following question. On one hand the proton as a quantum object is in a pure state and is described by a completely coherent wave function with zero entropy. On the other hand in high energy experiments (DIS) when probed by a small external probe, it behaves like an incoherent ensemble of (quasi-free) partons. Such an ensemble carries a nonvanishing ``classical'' entropy. Reference~\cite{Kharzeev:2017qzs} suggested that the origin of this entropy is entanglement between the degrees of freedom one observes in DIS (partons in the small spatial region of the proton) and the rest of the proton wave function which are not measured in the final state and therefore play the role of  an ``environment''. 

According to this idea, the lack of coherence and large entropy of the partonic density matrix which describes DIS within the parton model approach is due to entanglement of the observed partons with the unobserved proton degrees of freedom. If one knew the proton wave function, one would be able to calculate this density matrix by reducing it with respect to the unobserved ``environment''.
\begin{equation}
\hat\rho_{\rm PM}={\rm Tr}_{\rm unobs}\, \Big[\, |P\rangle\langle P| \, \Big]\,,
\end{equation}
where $|P\rangle$ is the proton wave function and the partial trace is taken over the unobserved degrees of freedom (the nature of which is not important at the moment). 
The entropy of the parton model is then identified with the von Neumann entropy of the reduced density matrix according to
\begin{equation}
S_{\rm PM}=-{\rm Tr} \Big[\, \hat\rho_{\rm PM}\, \ln \hat\rho_{\rm PM}\, \Big]\,.
\end{equation}
This proposal in principle eliminates the tension between the pure nature of the proton state and incoherent nature of the parton model. 

However a little thought shows that this is not the only way to eliminate this tension. 
The point is that the set of measurements that is described by the parton model is not complete, in the sense that it does not provide exhaustive information about the density matrix, even just about the density matrix of the observed degrees of freedom. In DIS the only quantity one measures is the average number of particles
\begin{equation}
	\langle N \rangle={\rm Tr} \Big[ \int \frac{d^2k}{(2\pi)^2} a^\dagger(\v{k}) a(\v{k})\, \hat\rho_{\rm PM} \Big]\,.
\end{equation}
Here, we suppress the longitudinal momentum label $x$ in order to illustrate our point in the simplest setting.
Extending to transverse momentum distributions (TMD's) one probes the average particle density at a fixed transverse momentum $\v{k} = (k_1,k_2)$: $\langle a^\dagger(\v{k}) a(\v{k}) \rangle$.
 Even considering more general measurements, such as those of double parton distributions, and possibly multi parton distributions one only probes the averages of the type $ \langle a^\dagger(\v{k}_1)a(\v{k}_1)...a^\dagger(\v{k}_n)a(\v{k}_n) \rangle $.
 
All of these observables are diagonal in the number operator basis, and  therefore in principle carry no information about nondiagonal elements of the density matrix in this basis.  Thus there is an infinite number of density matrices which are completely equivalent for the limited purpose of describing the results of only these measurements.

Interestingly, this lack of knowledge of the actual density matrix of the system can be characterized by an entropy. We will dub this entropy ``{\it the entropy of ignorance}''. Consider the situation in which one in principle can only measure a defining set of observables $\{O_i\}$ which is not complete, i.e. does not include all coordinates and/or conjugate momenta of the given quantum system. A density matrix that reproduces the results of this set of measurements $\hat\rho(\alpha_j)$ is parametrized by some parameters $\alpha_j$, which loosely speaking correspond to possible values of the observables not included in the set $\{O_i\}$.
 To each such density matrix one associates  von Neumann entropy
 \begin{equation}
 S(\alpha)=-{\rm Tr} \Big[\hat\rho(\alpha) \ln\hat\rho(\alpha)\Big]\,.
 \end{equation}
 We define the entropy of ignorance as the maximum of $S(\alpha)$ with respect to variation of $\alpha$
 \begin{equation}
 S_I=-{\rm Tr}\Big[\hat\rho(\bar\alpha)\ln\hat\rho(\bar\alpha)\Big]; \quad \quad \bar\alpha:\quad \frac{\partial}{\partial\alpha_j}S(\alpha)|_{\bar\alpha}=0\,.
 \end{equation}
 In Appendix A we give some examples of $S_I$ and its dependence on the defining set of observables in a simple quantum mechanical model.
 
 In the case of parton model the set $\{O_i\}$ includes all powers and products of the particle density operators $a^\dagger(k)a(k)$. Thus only diagonal matrix elements of the density matrix written in the Fock (particle number) basis are determined by the defining set of observables.  The parameters $\alpha_j$ therefore parametrize the off diagonal matrix elements of $\hat\rho$ in the particle number basis. The parameters $\bar\alpha$ defining the entropy of ignorance correspond to diagonal $\hat\rho$. 
 In order to prove this (see Refs.~\cite{Wehrl:1978zz,Witten:2018zva} for details) 
 consider $\hat \rho (t) = (1-t) \hat \rho_D  + t \hat \rho$, where $\rho_D$ is obtained from $\hat\rho$ by  dropping the off-diagonal elements of the density matrix.   
 Owing to the normalization condition, the variation of the entropy with respect to a parameter 
 \begin{equation}
	 \frac{\partial S}{\partial t}  = - \Tr  \left[ \frac { \partial \hat \rho(t)  }{ \partial t} \ln \hat \rho(t) \right].  
 \end{equation}
 Therefore at $t=0$  
 \begin{equation}
	 \left. \frac{ \partial S } {\partial t} \right|_{t=0}   = 
	 - \Tr \left[ \left( \hat\rho - \hat \rho_D \right)\ln \hat \rho_D \right]  = 0 .
 \end{equation}
 Then due to the concavity of the von Neumann entropy 
 \begin{equation}
	 \frac{ \partial^2 S } {\partial t^2} \le 0,   
 \end{equation}
 one concludes that $S(\rho(t=1))  \le  S(\rho(t=0))$ or in other words  
 \begin{equation}
	 S \le S_I. 
 \end{equation}
 %This is intuitively very clear, and indeed can be proven rigorously, see for example  Sect. 3.3. of Ref.~\cite{Witten:2018zva}.

Interestingly, since the matrix $\hat\rho(\bar\alpha)$ is diagonal in the particle basis, the entropy of ignorance is exactly equal to the Boltzmann entropy of the classical ensemble of partons with the probability distribution where probability to find  the system with $n$ particles is equal to the corresponding diagonal matrix element of $\hat\rho(\bar\alpha)$
\begin{equation}
S_I=S_B=-\sum_n\, p_n\ln p_n; \quad \quad p_n= \langle n|\hat\rho|n \rangle \,.
\end{equation}

Note that in this particular case, i.e. when the defining set of observables is a complete set of operators diagonal in a particular basis, the entropy of ignorance becomes identical with the so-called diagonal entropy introduced and studied 
 in Refs.~\cite{Barankov:2008qq,Santos_2012,PhysRevLett.107.040601}. This quantity is  defined as
 $S_D=-\sum_i\ p_i\ln p_i$, where $p_i$'s are diagonal matrix elements of the  density matrix in a specific basis. In \cite{Barankov:2008qq,Santos_2012,PhysRevLett.107.040601} the authors where primarily concerned with understanding  of the nature of equilibration (thermalization)  and thus considered  diagonal entropy in the energy basis. Our physical motivation here is different, however the formal similarity of the two quantities is interesting and may be useful to explore in future.

 An amusing case for the entropy of ignorance arises if we consider a system in a pure state. In this case, the von Neuman entropy is strictly zero; however if we ignore the off-diagonal elements of the density matrix and compute the entropy of ignorance the result is non-zero. We will consider this interesting situation in the context of our model wave function below.   

Since the classical parton model entropy is given by the entropy of ignorance, this begs the question whether the entanglement entropy in the sense of Ref.~\cite{Kharzeev:2017qzs} plays any role in the physics of parton model, or at the very least is not too different from the entropy of ignorance. Our goal in this paper is to compare the entanglement entropy and the ignorance entropy in a computable  model which has been used in recent years in the context of high energy scattering - the Color Glass Condensate (CGC) model.

The outline of this paper is as follows. In Section II we describe the CGC wave function. We point to certain similarity between reducing the CGC density matrix over the valence degrees of freedom and reducing the proton density matrix with respect to the ``environment'' alluded to earlier. In Section III we consider the Renyi entropy. We calculate the Renyi entropy of entanglement and the Renyi entropy of ignorance and compare the two. We find that the contribution of very high transverse momentum modes to the two entropies is the same to leading power in $1/k^2$, but the contribution of modes with momentum equal or smaller than the saturation momentum differs by a factor of order one. In Section IV we extend the discussion to the von Neumann entropy. Here we find that the discrepancy between the entanglement and ignorance entropies at high momenta is somewhat more significant. For large $k$ modes the two are still equal, but the relative difference between the two vanishes as a power of momentum enhanced by a power of the logarithm. At low momentum, however the relative difference between the two von Neumann  entropies is the same as between the two Renyi entropies. In Section V we consider the entropy of ignorance, but this time for a fixed configuration of the valence color charge density, We find that even for a fixed typical configuration of the valence fields the ignorance entropy approximates well the Boltzmann entropy of the partons, whereas the entanglement entropy in this case is strictly zero.
Finally we close with a discussion in Section VI.

\section{The CGC wave function}

%{\bf ALEX, we need to cite something here eg Ref \cite{Kovner:2007zu,Altinoluk:2009je} } 

We now introduce the CGC wave function \cite{Kovner:2007zu,Altinoluk:2009je} that we will use in our calculation.  
   
The Color Glass Condensate describes scattering at high energy.
For an ultra relativistic hadron,  large fraction of momentum is carried by the valence quarks and gluons. Due to their quantum nature, partons carrying large fraction of momentum radiate low energy gluons which have a lifetime relatively short to that of the valence charges. To put it in another way, the valence (``hard'') partons can be treated as static sources of the soft gluons. 

%It is natural to think how much  correlation between hard and soft parts in the proton wave function. One of the quantities to measure the correlation strength is entanglement entropy. 
The wave function of the system of slowly evolving valence charges and faster soft gluon degrees of freedom 
has the form 
\begin{equation}
| \psi \rangle = | s \rangle  \otimes   | v \rangle  \,,
\end{equation}
where $| v \rangle $ is the  state vector characterizing the valence degrees of freedom and  $| s \rangle$ is the vacuum of the soft fields in the presence of the valence source. Despite appearances, the state is not of a direct product form since the soft vacuum depends on the valence degrees of freedom.

In the leading perturbative order the CGC soft vacuum has the form
\begin{equation}
    | s \rangle=\mathcal{C}| 0\rangle
\end{equation}
with the coherent operator
\begin{equation}
	\label{Eq:CO}
	{\cal C}=\exp\left\{2 i {\rm tr} \int_{\v{k}} b^i(\v{k}) \phi^a_i(\v{k}) \right\}\,,
\end{equation}
where 
\begin{equation}
\phi_i(\v{k})\equiv a_i^+(\v{k})+a_i(-\v{k})\ ,
\end{equation}
the trace is  over all colors and the transverse vector is denoted by $\v{k} = (k_1,k_2)$. 
We use the following notation 
\begin{equation}
\int_{\v{k}} = \int \frac{d^2k }{(2\pi)^2} \,.
\end{equation}
The background field $b^i_a$ is determined by the valence color charge density $\rho$ via:
\begin{equation}
    b^i_a(\v{k})=g\rho_a(\v{k})\frac{i\v{k}_i}{k^2} + c_a^i(\v{k})\,.
	\label{Eq:b}
\end{equation}
The correction $c_a^i(\v{k})$ is suppressed by at least ${\cal O}(\rho^2)$ at small charge density, and we will neglect it in the following. It can be taken  into account as a perturbation, but we believe our results are stable to this particular correction. Note also that $c^i$ is transverse, that is $\v{c} \cdot \v{k} = 0$. Therefore at the leading order in $\rho(\v{k})$, only gluons with the longitudinal  polarization contribute to ${\cal C}$ and $  | s \rangle$. 

The valence wave function $|v\rangle$ is customarily modeled in the so called McLerran-Venugopalan (MV) model as~\cite{McLerran:1993ni,McLerran:1993ka}
\begin{equation}\label{mv}
  \langle \rho | v \rangle \langle v |  \rho \rangle=   {\mathcal N}e^{-\int_{\v{k}}
 \frac{1}{2\mu^2} \rho_a(\v{k})   
  \rho^*_a(\v{k})}\,,
\end{equation}
where ${\mathcal N}$ is the normalization factor and the parameter $\mu^2$ determines the average color charge density in the valence wave function.
Note that Eq.~(\ref{mv}) does not determine the (possibly $\rho$-dependent) phase of $|v\rangle$. This phase however does not enter our calculation.
 
Consider the hadron density matrix:
\begin{equation}
    \hat{\rho}=|v\rangle\otimes  |s\rangle \langle s| \otimes \langle v| \,.
\end{equation}
In the following we will integrate out the valence (slow) degrees of freedom  and derive the reduced density matrix for the soft gluons. That is we compute the reduced 
density matrix 
\begin{equation}
    \hat{\rho}_r=  {\rm Tr}_\rho \hat\rho\equiv  \int D \rho\,  \langle \rho | \hat{\rho}  |  \rho \rangle =  \int D \rho\,  \langle \rho |v\rangle\,   |s\rangle \langle s|    \, \langle    v|  \rho \rangle\,.
\end{equation}
We will then use this density matrix for calculating the entanglement entropy of the soft gluons and compare it to the entropy of ignorance. 

We expect this  model to be a meaningful proxy to study the question discussed in the introduction. One obvious common element between our model calculation and the real life parton model is the natural bi-partitioning of the degrees of freedom in the underlying wave function and integrating over  the ``environment''. Physically though the analogy goes a little further. In our model approach we will be reducing the density matrix over the slow degrees of freedom.  The parton model in QCD has a similar meaning. At large transverse momentum ($Q^2$) the observed partons correspond to the faster degrees of freedom. The unobserved ``environment'' that has to be integrated out presumably consists of lower transverse momentum modes (or in coordinate space modes extending outside the spatial region probed by the virtual photon) which have lower frequency than the high transverse momentum partons, and possibly confinement scale nonperturbative glue which again naturally has much lower frequencies. Thus, although the analogy may not be perfect, we believe that our toy model captures some basic relevant physics  and therefore can teach us a meaningful lesson about the actual QCD parton model.

\section{Density matrix in number representation and the Renyi entropy}

Using the MV model for the valence degrees of freedom, the reduced density matrix is calculated as
\begin{align}
    \hat{\rho}_r=\mathcal{N}\int D \rho\, \,  e^{-\int_{\v{k}}\frac{1}{2\mu^2}\rho_a(\v{k})\rho^*_a(\v{k})}\mathcal{C}(\rho_b,\phi_b^i)|0\rangle \langle0|\mathcal{C}^\dagger(\rho_c,\phi_c^j) \, . 
\end{align}
The very same reduced density matrix was obtained, and the von Neumann entropy was calculated in previous papers of some of the authors \cite{Kovner:2015hga,Kovner:2018rbf,Armesto:2019mna}. The calculation was performed in the field basis. Since the gluon number basis plays a special role in our current discussion, we will perform this calculation independently using this basis. Here because of the particularity of Eq.~\eqref{Eq:b} in the leading order, we consider longitudinally and transversely polarized gluons with corresponding annihilation operators defined as 
$a_c^\|(\v{k})   = \v{k} \cdot \v{a}_c (\v{k}) / |\v{k}|$ and  
$a_c^\perp(\v{k})   = \epsilon_{ij} k^i a^j_c (\v{k}) / |\v{k}|$.

%The goal of the current section is to extract diagonal components of this density matrix in gluon number basis, and compare with the full result at $\Tr(\rho^2)$ level.

We label the basis states as 
\begin{align}
\prod_{c}\prod_{\lambda}\prod_{k}|n_c^\lambda(\v{k}), m_c^\lambda(-\v{k})\rangle=\prod_{c}\prod_{\lambda}\prod_{k}|N_{c}^{\lambda}(\v{k})\rangle
\,,
\label{Eq:Doublers}
\end{align}
where $\lambda = \|, \perp$ and $c$ are the polarization and color indices respectively.
We have introduced for convenience $N_c^i=n^i_c+m^i_c$. The reason for introducing Eq.~\eqref{Eq:Doublers} is that in our density matrix, a
mode with momentum $\v{k}$ mixes only with the mode with momentum $-\v{k}$ due to the fact that  $\rho^*_a(\v{k})=\rho_a(-\v{k})$. In addition the density matrix 
is translationally invariant, which has a consequence that $\hat\rho_r$ is a direct product of density matrices in a fixed transverse momentum sector. 

 The continuum states are customarily normalized as,
\begin{align}
 \langle n_c^\lambda(\v{k})|n_c^\lambda(\v{k}')\rangle=\langle 0|\frac{[a^\lambda_c(\v{k})]^n}{\sqrt{n!}} \frac{[a^{\lambda\dagger}_c(\v{k}')]^n}{\sqrt{n!}}|0\rangle
\end{align}
with the corresponding orthogonality relation 
\begin{align}
	\langle \v{k} | \v{k}'\rangle=\langle 0| a_c^\lambda(\v{k}) a_{c'}^{\lambda' \dagger}(\v{k}')|0\rangle=(2\pi)^2 \delta_{\lambda \lambda'}  \delta_{c c'} \delta^2(\v{k}-\v{k}')\,.
\end{align}
For convenience we discretize momentum by putting the system inside the spatial region of area $S_\perp$ and granularity $\Delta$.
Then
\begin{align}
   \langle \v{k}|\v{k}'\rangle=\frac{(2\pi)^2 }{\Delta^2}  \delta_{\lambda \lambda'}  \delta_{c c'}  \delta_{kk'}
\end{align}
with $S_\perp \Delta^2 = (2\pi)^2$. 
We also find it easier to work with the states which have a unit norm, as this makes the interpretation of diagonal matrix elements as probabilities straightforward. We thus redefine the multi gluon states as 
\begin{align}
 \prod_{c}\prod_{\lambda }\prod_{k}|n_c^\lambda( \v{k}), m_c^\lambda(-\v{k})\rangle=\prod_{c}\prod_{\lambda}\prod_{k} \left(\frac{[a^{\lambda\dagger}_c(\v{k}){\frac{\Delta}{2\pi}}]^{n_c^\lambda}}{\sqrt{n_c^\lambda!}} \right) \left(\frac{[a^{\lambda\dagger}_c(-\v{k}){\frac{\Delta}{2\pi}}]^{m_c^\lambda}}{\sqrt{m_c^\lambda!}}\right)|0\rangle\,
\end{align}
and use this normalization in the rest of the paper.

\subsection{Entropy of entanglement}
From the structure of the density matrix it is obvious that it is a direct product over color. We thus consider the calculation for a fixed color index $c$.
%Since the final result of any matrix element is a product of contribution from different colors. It is possible that, for some color, it's diagonal and not for other colors. We start discussing the structure of off-diagonal terms for one color. At this point, we wonder what's the order of magnitude of  off-diagonal contribution compared to diagonal part.

The action of the coherent operator on the soft gluon vacuum can be represented as
\begin{align}
    {\cal C}|0\rangle&=e^{i \int_{\v{k}}  b^i_c(\v{k})  [{a^i_c}^+(\v{k})+a^i_c(-\v{k})]}|0\rangle
    %&=e^{-\int_{\v{k}} \frac{gk_i}{k^2}\rho(\v{k})a_i^+(\v{k})- \frac{gk_i}{k^2}\rho(-\v{k})a_i(\v{k})}|0\rangle\\
    %&=e^{-\int_{\v{k}} \frac{gk_i}{k^2}\rho(\v{k})a_i^+(\v{k})}e^{\int_{\v{k}}\frac{gk_i}{k^2}\rho(-\v{k})a_i(\v{k})}e^{\frac{1}{2}\int_{\v{k}} \int_{\v{k'}}\sum_i \frac{g^2k_ik_i'}{k^4}[a_i^+(\v{k}),a_i(\v{k'})]}|0\rangle\\
    %&=e^{-\int_{\v{k}} \frac{gk_i}{k^2}\rho(\v{k})a_i^+(\v{k})}e^{-\frac{1}{2}\int_{\v{k}}\frac{g^2}{  k^2}|\rho(\v{k})|^2}\\
    =e^{i\int_{\v{k}} b^i_c(\v{k}) {a^i_c}^+(\v{k})}  
       e^{-\frac{1}{2} \int_{\v{k}} \frac{g^2}{k^2}|\rho_c(\v{k})|^2} |0\rangle\,,
\end{align}
where we used Baker--Campbell--Hausdorff formula. 
%The complex conjugate of the above result would correspond to ${\cal C}^{\dagger}$. 
We can then write $\hat{\rho}_r$ as
\begin{align}
    \hat{\rho}_r  &={\mathcal N}\int  \prod_{\v{k} } \prod_a  d\rho_a(\v{k})\, \,  
    e^{-\frac{\Delta^2}{(2\pi)^2}
    \left(\frac{1}{2\mu^2}+\frac{g^2}{k^2}\right)
    \rho_a(\v{k})\rho^*_a(\v{k})}     
    e^{ib_a^i(\v{k})a_{i a}^{\dagger}(\v{k})\frac{\Delta^2}{(2\pi)^2}}   
    |0\rangle \langle0|
    e^{-ib_a^{*i}(\v{k})a_{ia}(\v{k})   \frac{\Delta^2}{(2\pi)^2}} \notag \\
    &={\mathcal N}\int  \prod_{\v{k} \geq 0} \prod_a  d\rho_a(\v{k})d\rho_a(-\v{k})\, \, 
     e^{-2\frac{\Delta^2}{(2\pi)^2}\left(\frac{1}{2\mu^2}+\frac{g^2}{k^2}\right)\rho_a(\v{k})\rho^*_a(\v{k})}
     e^{i\frac{\Delta^2}{(2\pi)^2}\left(b_a^i(\v{k})a_{ia}^{\dagger}+b_a^{*i}(\v{k})a_{ia}^{\dagger}(-\v{k})\right)}
     |0\rangle \notag  \\
      &\langle0|e^{-i\frac{\Delta^2}{(2\pi)^2} \left(b_a^{*i}(\v{k})a_{ia}(\v{k})+b_a^{i}(\v{k})a_{ia}(-\v{k})\right)} \,.
      \label{Eq:rho_prelim}
\end{align}

Consider the matrix element $\prod_{\lambda} \langle n_c^\lambda(\v{q}), m_c^\lambda(-\v{q}) |\hat{\rho}_r(\v{q})|\alpha_c^\lambda(\v{q}), \beta_c^\lambda(-\v{q})\rangle$.  
 Since all operators inside exponential  commute with each other in Eq.~\eqref{Eq:rho_prelim}, we have
 \begin{align}
     &\langle n_c^\|(\v{q}), m_c^\|(-\v{q}) |e^{i  \frac{\Delta^2 }{(2\pi)^2} (b_a^i(\v{k})a_{ia}^{\dagger}+b_a^{*i}(\v{k})a_{ia}^{\dagger}(-\v{k}))}|0\rangle \notag \\
     =&\langle n_c^\|(\v{q}), m_c^\|(-\v{q}) | 
     \prod_b \prod_t \sum_{n^t,m^t} \frac{\left(i\frac{\Delta^2}{(2\pi)^2}b_b^{t}(\v{q})a_{tb}^{\dagger}(\v{q})\right)^{n^t}}{n^t!}
     \frac{\left(i\frac{\Delta^2}{(2\pi)^2}b_b^{*t}(\v{q})a_{tb}^{\dagger}(-\v{q}) \right)^{m^t}}{m^t!}
     |0\rangle \notag  \\
     =&\langle n_c^\|(\v{q}), m_c^\|(-\v{q}) |
     \frac{ \left( i\frac{\Delta^2}{(2\pi)^2}b_c^{\| }(\v{q})a_{\| c}^{\dagger}(\v{q}) \right)^{n^\|}}{n^\|!}
     \frac{\left( i\frac{\Delta^2}{(2\pi)^2}b_c^{*\|}(\v{q})a_{\| c}^{\dagger}(-\v{q}) \right)^{m^\|}}{m^\|!}
     |0\rangle \notag  \\
     =&(i)^{n^\|+m^\|}\left(\frac{\Delta^2}{(2\pi)^2} \right)^{\frac{n_\|+m_\|}{2}}\frac{(b_c^\| (\v{q}))^{n^\|}}{\sqrt{n^\|!}}\frac{(b_c ^{*\|} (\v{q}))^{m^\|}}{\sqrt{m^\|!}}
     =
     (-1)^{n^\|}\left(\frac{ g}{q}\right)^{n^\|+m^\|}
     \left(\frac{\Delta^2}{(2\pi)^2}\right)^{\frac{n_\|+m_\|}{2}}
     \frac{[\rho_c(\v{q}) ]^{n^\|}}{\sqrt{n^\|!}}\frac{[\rho_c(-\v{q})]^{m^\|}}{\sqrt{m^\|!}}\,\end{align}
and the trivial  
 \begin{align}
	 &\langle n_c^\perp(\v{q}), m_c^\perp|(-\v{q}) |e^{i  \frac{\Delta^2 }{(2\pi)^2} (b_a^i(\v{k})a_{ia}^{\dagger}+b_a^{*i}(\v{k})a_{ia}^{\dagger}(-\v{k}))}|0\rangle  = 
	 \delta_{ n_c^\perp, 0 } 
	 \delta_{ m_c^\perp, 0 }\,. 
 \end{align}
 The latter indicates that the gluons with the transverse polarization contribute only to partonic vacuum; they are in the pure state and thus do no contribute to entropy. We will thus consider only longitudinally polarized gluons.   
 Integration with respect to $\rho_a(\pm\v{k})$  can now be carried out 
 in Eq.~\eqref{Eq:rho_prelim}.  For the integral to yield  a non-zero value it is required that  
 \begin{equation}
  %\sum_s \left( 
  n_\|+\beta_\| %\right) 
  =%\sum_s 
  %\left(
  m_\|+\alpha_\|. %\right). 
 \end{equation}
Thus the required matrix element is 
\begin{align}
     &\langle n_c(\v{q}), m_c(-\v{q}) |\hat{\rho}_r(\v{q})|\alpha_c(\v{q}), \beta_c(-\v{q})\rangle 
     = \mathcal{N} \left[2\frac{\Delta^2}{(2\pi)^2}\left(\frac{1}{2\mu^2}
     +\frac{g^2}{q^2}\right)\right]^{ -n-\beta-1}
      \left(\frac{g^2}{ q^2}\frac{\Delta^2}{(2\pi)^2}\right)^{ n+\beta}
            %\notag \\ &\times
      \frac{
      \left( n+\beta\right)!
      }{\sqrt{n!m!\alpha!\beta!}}
	  \delta_{n+\beta, m+\alpha}\,,
\end{align}
where we left out the  polarization label, as only $\|$ contributes to the non-trivial part of the density matrix. 

To calculate the Renyi entropy we need to find ${\rm Tr}  \ \hat \rho_r^2$. This requires  squaring   the matrix element 
and summing with respect to all possible $n$, $m$, $\alpha$, $\beta$. 
Most efficiently this can be done by using an integral representation for the factorial $\left(n+\beta\right)!$: 
 \begin{equation}
 \left(n+\beta\right)!  = \int_0^\infty dt_1 t_1^{ \,n+\beta } e^{-t_1}
 \end{equation}
 and for the Kronecker delta function 
  \begin{equation}
  \delta_{\left( n+\beta \right), \left(m+\alpha\right)}
   = \frac{1}{2\pi i}  \oint_{\cal C}  \frac{dz}{z} z^{ \left( n+\beta - m-\alpha \right) }\, ,  
 \end{equation}
 where ${\cal C}$ is a unit circle.  
 The normalization $\mathcal{N}$ is fixed by requiring that 
${\rm Tr} \ \hat \rho_r = 1$. 
This leads to
\begin{equation}
\mathcal{N} = 2\frac{\Delta^2}{(2\pi)^2}\left(\frac{1}{2\mu^2}+\frac{g^2}{q^2}\right)
     \left(1-R\right)\, , 
\end{equation}  
  where 
 \begin{equation}
 R =\left(1+\frac{q^2}{2g^2 \mu^2} \right)^{-1}\,.
 \end{equation}

The final expression for the matrix element including the normalization is: 
\begin{align}\label{matel}
     & \langle n_c(\v{q}), m_c(-\v{q}) |\hat{\rho}_r(\v{q})|\alpha_c(\v{q}), \beta_c(-\v{q})\rangle 
	 = (1-R) 
	           % \notag \\ &\times
      \frac{
      \left( n+\beta\right)!
  }{\sqrt{n!m!\alpha!\beta!}}\left( \frac {R}{2} \right)^{n+\beta} 
      \delta_{ \left( n+\beta \right), \left(m+\alpha\right)}\,. 
\end{align}

For the trace of the square of the density matrix we get
    \begin{align}\label{matels}
     &\sum_{m,n,\alpha,\beta}\Big(\prod_{s} \langle n_c(\v{q}), m_c(-\v{q}) |\hat{\rho}_r(\v{q})|\alpha_c(\v{q}), \beta_c(-\v{q})\rangle\Big)^2\notag \\
     =&
    % \mathcal{N}^2 \left[2\frac{\Delta^2}{(2\pi)^2}\left(\frac{1}{2\mu^2}+\frac{g^2}{q^2}\right)\right]^{-2} 
	(1-R)^2 
     \frac{1}{2\pi i}  \oint  \frac{dz}{z}  
     \int dt_1 d t_2 
     e^{-t_1 -t_2}
     \notag \\ &\times
     \sum_{m, n, \alpha, \beta}  
    \frac{ 1   } { n!m!\alpha!\beta! } 
     \left[ 
     t_1  z^{-1} \frac{ R }{2} 
     \right]^{m} 
     \left[ 
      t_2 z \frac{ R }{2}  
     \right]^{n} 
     \left[ 
     t_1  z^{-1} \frac{R }{2} 
 \right]^{\alpha }
     \left[ 
      t_2 z \frac{ R }{2} 
     \right]^{\beta} =%\mathcal{N}^2 \left[2\frac{\Delta^2}{(2\pi)^2}\left(\frac{1}{2\mu^2}+\frac{g^2}{q^2}\right)\right]^{-2} 
	 \frac{(1-R)^2}{1-R^2}\,
 \end{align}
and  the final result for ${\rm Tr} \, \hat \rho_r^2$
\begin{equation}
\sum_{m,n,\alpha,\beta}\Big( \langle n_c(\v{q}), m_c(-\v{q}) |\hat{\rho}_r(\v{q})|\alpha_c(\v{q}), \beta_c(-\v{q})\rangle\Big)^2 = 
    \frac{1-R}{1+R}  = \frac{1}{1+4 \frac{g^2 \mu^2}{q^2}}\,.
\end{equation}  
At small momentum this ratio goes to zero, and at large momentum it approaches unity. 

The Renyi entropy is thus
\begin{align}
    S_R = -\ln \Tr\,  \hat \rho_r^2 =
    \frac{1}{2} (N_c^2-1) S_\perp\int \frac{d^2q}{(2\pi)^2}
    \ln(1+4\frac{g^2 \mu^2}{q^2} )\,. 
\end{align}
The color factor arises since the density matrix is a product of density matrices over the color index, while the area factor appears due to taking the continuum limit in the sum over momentum,

This coincides with result obtained in Ref.~\cite{Kovner:2015hga}. In number representation basis, we were thus able to reproduce the result of the previous calculations of the entanglement  entropy which were performed in the field basis.

\subsection{Entropy of ignorance}
We now turn to the calculation of the entropy of ignorance. To do that, as discussed above we replace $\hat\rho_r$ by only its diagonal part in the gluon number basis, $\hat\rho_I$.
%Now we will approximate the full density matrix by only its diagonal elements $\hat \rho_i$. 
%In this case the normalization of the matrix does not change and additionally it remains to be positive definite. 
%Since the off-diagonal terms are neglected, we expect the entropy in this case to be larger than the entropy of the full reduced matrix. 

 Then diagonal matrix elements of the density matrix for a given value of momentum $q$ are 
 \begin{align}
 & \prod_{c}\langle n_c(\v{q}), m_c(-\v{q}) |\hat{\rho}_I(\v{q})|n_c(\v{q}), m_c(-\v{q})\rangle
 %notag \\
 =
 (1-R) 
	           % \notag \\ &\times
      \frac{
      \left( n+m\right)!
  }{n!m!}\left( \frac {R}{2} \right)^{n+m} \,. 
 \end{align}

 For  $\tr(\rho_I^2)$ at fixed momentum and color index we evaluate the following 
\begin{align}
\tr(\rho_I^2) =  
 (1-R)^2 
	           % \notag \\ &\times
			   \sum_{m,n} \left[ \frac{
      \left( n+m\right)!
  }{n!m!}\left( \frac {R}{2} \right)^{n+m} \right]^2
= \frac{  (1-R)^2 }{\sqrt{1-R^2}}\, . 
\end{align}
where  the sum is computed in App.~\ref{App}. 
%
%\begin{equation}
% \mathfrak{S} = 4  \int_0^\infty dx \, x  
%     K_0(2x) I_0^2\left(\frac{ q_1^2}{q^2}  {R} \, x \right)  I_0^2\left(\frac{ q_2^2}{q^2}  {R}\,  x \right)\,.
%\end{equation}   

The associated Renyi entropy is given by 
\begin{align}
    S_R^I= - \ln \Tr \rho^2_I = 
    \frac{1}{2} (N_c^2-1) S_\perp\int \frac{d^2q}{(2\pi)^2}
    \ln
	\left[\left(1+2 \frac{g^2 \mu^2}{q^2}\right) \sqrt{ 1 + \frac{4 g^2\mu^2}{q^2 } } \right]\,.
\end{align}

The two expressions $S_R$ and $S_R^I$ are clearly different. They do coincide however in the limit of high transverse momentum. Considering  the contribution from high momenta $q^2\gg g^2\mu^2$, we find
\begin{equation}\label{leadingR}
S_R^I(q^2)_{q^2\gg g^2\mu^2}\approx \frac{1}{2}(N_c^2-1)S_\perp\frac{4g^2\mu^2}{q^2}\approx S_R(q^2)_{q^2\gg g^2\mu^2}\,.
\end{equation}
Thus the leading contribution of the high momentum modes to the ignorance and entanglement entropies is the same.
The first sub-leading term is different
\begin{align}\label{subleadingR}
  [S_R^I(q^2)-S_R(q^2)]_{q^2\gg g^2\mu^2}   \approx
    (N_c^2-1) S_\perp%\int 	\frac{d^2q}{(2\pi)^2}
    %\frac{4 q^4-2(q_1^4 + q_2^4)}{q^8} g^4 \mu^4 \,,
	\left( \frac{  g^2 \mu^2  } { q^2 } \right)^2\,. 
\end{align}
 We will discuss this feature in the last section.
 
\begin{figure}[t]
	\centering 
	\includegraphics[width=0.45\textwidth]{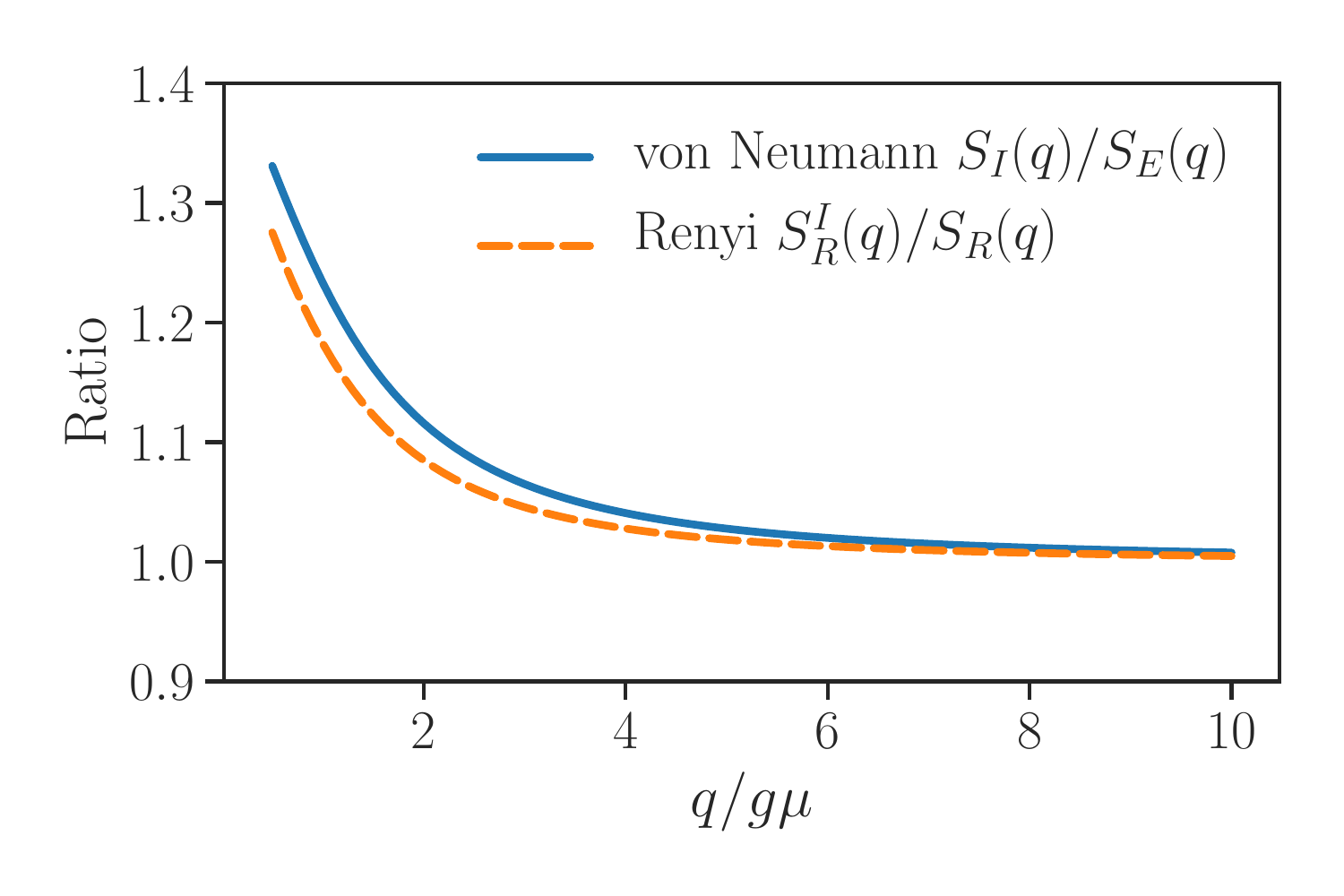}
	\caption{Ratios of entropy densities at a given magnitude of the transverse momentum $q/g\mu$. $S_I(q^2)$ is the von Neumann entropy density of ignorance and  $S_E(q^2)$ is the corresponding entanglement entropy density. The same for Renyi entropy densities.}
	\label{fig:Ratio}
\end{figure}

 At momenta of order $g\mu$ and smaller, i.e. in the saturation regime, the two entropies are substantially different. The ratio between  the two is plotted on Fig.~\ref{fig:Ratio}. At zero momentum the ratio depicted in Fig.~\ref{fig:Ratio} tends to $3/2$, since $S_R ({q^2\rightarrow 0}) \sim\ln 1/q^2$ while $S_R^I ({q^2\rightarrow 0})\sim \ln 1/q^3$.
%At moderate and small momenta the difference between the two is rather substantial.
%One can compute numerically~\footnote{Find Mathematica code in Ref.~\cite{Skokov2019}.}  the difference between the two due to momentum modes
% $q\le g \mu$
%\begin{align}
%    S_R^I-S_R =  
%    \frac{1}{2} (N_c^2-1) S_\perp\int_{q^2<g^2\mu^2} \frac{d^2q}{(2\pi)^2}
%    \ln
%    \left[\frac{\left(1+2 \frac{g^2 \mu^2}{q^2}\right)^2 } { \left(1+4 \frac{g^2 \mu^2}{q^2}\right)}   \mathfrak{S}^{-1}\right]\,.
%\end{align}
%At large momentum 
%\begin{align}
%    S_I-S_R   \approx
%    \frac{1}{2} (N_c^2-1) S_\perp\int 	\frac{d^2q}{(2\pi)^2}
%    \frac{q_1^4 + q_2^4}{q^8} g^4 \mu^4 
%\end{align}
%One finds   $(S_R^I-S_R)/S_R \approx 0.36$. 
%That is $S_R$ underestimates  $S_R^I$ by about 40\%.   

\section{von Neumann entropy} 
\label{Sect:vN}

Let us now study the behavior of the von Neumann entropy.
% One of the main differences between the Renyi and von Neumann entropies is that the leading contribution to the Renyi entropy is dominated by the ``vacuum'' contribution (that is with zero parton number) at large transverse momentum; the corrections are suppressed by $1/k^2$. The vacuum contribution is identical for the entanglement and ignorance entropies; thus it is not surprising that the difference between them is negligible at large $|\v{k}|$. This ``accidental'' masking of the difference motivated us to consider the von Neumann entropy, for which we expect to get somewhat distinct result. 

\subsection{Entropy of entanglement}
The  entanglement entropy in this model was calculated  in Ref.~\cite{Kovner:2015hga}. The complete final result (adjusting for a different normalization of $\mu^2$ used in Ref.~\cite{Kovner:2015hga}) is
\begin{align}
    S_E=\frac{1}{2} (N_c^2-1) S_\perp\int \frac{d^2q}{(2\pi)^2} \Bigg[\ln(\frac{g^2\mu^2}{q^2})+ \sqrt{1+4\frac{g^2\mu^2}{q^2}}  \ln (1+\frac{q^2}{2 g^2\mu^2}+ \frac{q^2}{2 g^2\mu^2}\sqrt{1+4\frac{g^2\mu^2}{q^2}} )    \Bigg]\,.
\end{align}
%The integrand approaches zero at large transverse momentum. 

%Use the r as magnitude of momentum k in units of $g\mu$, we have
%\begin{align}
%     S_C=\frac{1}{2} (N_c^2-1) (g^2\mu^2)S_\perp\int \frac{d^2r}{(2\pi)^2} \Big[\ln(\frac{1}{r^2})+ \sqrt{1+\frac{4}{r^2}}  \ln (1+\frac{r^2}{2}+ \frac{r^2}{2 }\sqrt{1+\frac{4}{r^2}})     \Big]
%\end{align}
\subsection{Entropy of ignorance}

The von Neumann entropy of ignorance for a single momentum mode $\v{q}$ is 
\begin{align}
  {S}_I(\v{q})=-\sum_{m,n} \rho_{n m} \ln \rho_{n m}
\end{align}
with 
\begin{align}
     &  \rho_{n m} = \langle n_c(\v{q}), m_c(-\v{q}) |\hat{\rho}_I(\v{q})|n_c(\v{q}), m_c(-\v{q})\rangle \, .
\end{align}
Supplementing the above by the integration with respect to the transverse momentum, the formal expression is
\begin{align}
    {S}_I&=-\frac{1}{2} (N_c^2-1) S_\perp \, 
	\int \frac{d^2 q}{(2\pi)^2}  
	\sum_{m ,n } 
	%\Bigg\{ 
		\left[
		(1-R) \frac{ (m+n)! } {m! n!}  \left( \frac{ R } 2 \right)^{m+n}   \right] \ln \left [
(1-R) \frac{ (m+n)! } {m! n!}  \left( \frac{ R } 2 \right)^{m+n} 
\right]%\Bigg\}
\,.
\end{align}
%The integral with respect to the angle can be performed following 
%\begin{align}
% &\int_0^{2\pi} d \theta \, \sin^{2m}(\theta)\cos^{2n}(\theta) \ln\big[A \sin^{2m}(\theta)\cos^{2n}(\theta)\big]=
%2  B\left(m+\frac{1}{2},n+\frac{1}{2}\right) \notag \\ &\times   \left(-(m+n) \psi(m+n+1)+m \psi\left(m+\frac{1}{2}\right)+n \psi\left(n+\frac{1}{2}\right) + \ln A\right)\,,
%\end{align}
%where $B(x,y)$ is the beta-function. 
%This leads to 
%\begin{align}
%     {S}_I&= - (N_c^2-1)  g^2\mu^2 S_\perp
%	\int \frac{r dr}{(2\pi)^2}  \sum_{m_{1,2},n_{1,2}} 
%	 \Bigg\{ 
	 %\sum_{m_1,m_2,n_1,n_2} 
%	 \left[
%	\frac{r^2 \Gamma(\sum_s (m_s+n_s) +1) }{ n_1! n_2! m_1! m_2!}
%	\left(\frac{1}{r^2+2}\right)^{\sum_s (m_s+n_s)+1}\right]\notag\\
%	&  \times  B\left(m_1+n_1+\frac{1}{2},m_2+n_2+\frac{1}{2}\right) 
%	\Bigg( \ln\left[\frac{r^2 \Gamma(\sum_s (m_s+n_s) +1)}{n_1!n_2!m_1!m_2!}\left(\frac{1}{r^2+2}\right)^{\sum_s (m_s+n_s)+1}\right]
%	\notag\\& + \sum_s (m_s+n_s) \psi \left(m_s+n_s+\frac{1}{2}\right)
%	 -\sum_s( m_s+n_s) \, \psi\left(\sum_s (m_s+n_s)+1\right)\Bigg) \Bigg\}\,.
%\end{align}
Unlike in the case of Renyi entropy we are unable to sum the series analytically. Numerically this can however be calculated; the resulting plot of the ratio of two entropies appears in Fig.~\ref{fig:Ratio}. We see that the differences for von Neumann entropy are somewhat more pronounced.

Just like for the Renyi case, we can study analytically the contribution of high momentum modes.
For large $q$  to the  sub-leading order we get  
\begin{align}\label{uvi}
% \frac{d 	\mathcal{S}_I }{r dr} 
  {S}_I(q)\simeq  
 %\int \frac{r d r}{(2\pi)^2}\, 
 \frac{ (N_c^2-1) g^2\mu^2 S_\perp  }{ %2 \pi
	 q^2} \left[ \ln  \left( e \frac{q^2}{g^2\mu^2} \right)  +\frac{g^2\mu^2}{q^2} 
 \ln \frac{e}{2} 
 \right] 
\end{align}
 and
\begin{align}\label{uve}
%	\frac{d    \mathcal{S}_E}{r dr}
{S}_E(q)	 \simeq    \frac{ (N_c^2-1) g^2\mu^2 S_\perp }{ %2\pi
 q^2} \left[  \ln \left( e \frac{q^2}{g^2\mu^2}  \right)  
 -  \frac{ g^2 \mu^2} {q^2 }  \ln \left( e  \frac{  q^4 } {   g^4 \mu^4} \right)  
 \right] \, .
\end{align}
Obviously, the leading behavior of the two expressions is the same. The subleading terms are different just like in case of the Renyi entropy. The difference is again a subleading power of $1/q^2$, but this time it is enhanced by $\ln q^2$.

At small momentum we find numerically that the ratio tends to 3/2 just like for the Renyi entropy.

  This larger discrepancy for von Neumann entropy is indeed demonstrated in Fig.~\ref{fig:Ratio}.

%  These expressions have been averaged over the angle. We have kept here the leading logarithmic term as well as the constant under the logarithm.
%We thus see that the leading logarithmic term is the same. So that the ratio ${S}_I/{S}_E$ goes to 1 from above at large momentum. This time, however, the relative difference between the two is of order $1/\ln q^2$ rather than $1/q^2$ as for the Renyi entropies.
%Indeed Fig.~\ref{fig:Ratio}  shows that the ratio stays significantly above one up to rather large values of momentum.

\section{Fixed color charge configuration} 
So far we have compared the entanglement entropy with the ignorance entropy of the reduced density matrix, which was obtained by tracing over the valence degrees of freedom. 
There is another instructive exercise we can do. Let us consider the density matrix for soft modes at a {\it fixed} configuration of the color charge density.  Recall that the valence charges are slow degrees of freedom, so that in any scattering event at high energy the valence charge density is fixed. So any given event essentially probes the hadronic wave function at fixed color charge distribution $\rho_a(\v{q})$. It is thus interesting to see how the entanglement and ignorance properties differ at fixed $\rho$.

As far as entanglement is concerned, the situation is completely trivial. At fixed $\rho_a(\v{q})$ the soft modes are in a pure state,
as can be easily seen from 
\begin{equation}
	\hat \rho  = {\cal C} |0\rangle \langle0|   {\cal C}^\dagger 
\end{equation} with a unitary ${\cal C}$, see Eq.~\eqref{Eq:CO}. 
Thus entanglement entropy at fixed $\rho$ strictly {\it vanishes}.

The ignorance entropy on the other hand
is not zero. Indeed, for a fixed configuration, the {\it diagonal} matrix element is 
\begin{align}
 \prod_{s} 
 \langle n_c(\v{q}), m_c(-\v{q}) |
 \hat{\rho}  (\v{q}) |n_c(\v{q}), m_c(-\v{q})
 \rangle  = 
 \frac{ 1} {n! m! } 
 e^{- 2  \frac{\Delta^2} { (2\pi)^2 } \frac{ g^2 } { q^2} | \rho_a (\v{q})|^2} 
 \left( \frac{ g^2 }  { q^2 }   
 \frac{\Delta^2 |\rho(\v{q})| ^4 }{(2\pi)^2}  \right)^{m +n}\,.
\end{align}
Therefore the associated  Renyi entropy is given by  
\begin{align}
{ S}_I = - \ln \Tr \hat {\rho}^2 
= \frac 12 S_\perp \int \frac{d^2q} { (2\pi)^2} 
 \sum_{a} 
\left[
4  \frac{g^2}{q^2} \frac{\Delta^2} {(2\pi)^2} 
 |\rho_a(q)|^2  - \ln I^2_0\left( \frac{2 g^2} {q^2} \frac{\Delta^2}{(2\pi)^2} |\rho_a(\v{q})|^2  \right) \right]\, .
\end{align}
A typical configuration in the MV model has the magnitude of order 
\begin{equation}
\frac{\Delta^2}{(2\pi)^2}|\rho_a(q)|^2\sim \mu^2\,.
\end{equation}
We thus obtain 
\begin{align}
	{ S}^{\rm typ}_I = - \ln \Tr \hat {\rho}^2 
= \frac 12 (N_c^2-1) S_\perp \int \frac{d^2q} { (2\pi)^2} 
\left[
4  \frac{g^2 \mu^2}{q^2}  - \ln I^2_0\left( \frac{2 g^2 \mu^2} {q^2} \right) \right]\, .
\end{align}
At hight momentum the integrand behaves as $4  \frac{g^2 \mu^2}{q^2}   -  2  \left( \frac{g^2 \mu^2}{q^2} \right)^2 $; compare this with 
%the entanglement entropy 
% $4  \frac{g^2 \mu^2}{q^2}   -  8  \left( \frac{g^2 \mu^2}{q^2} \right)^2 $ and 
the ignorance entropy   $4  \frac{g^2 \mu^2}{q^2}   -  6  \left( \frac{g^2 \mu^2}{q^2} \right)^2$  of the reduced density matrix.

That is if we fix a typical configuration of the color charges $\rho_a(\v{q})$, the ignorance entropy we obtain is very close to the ignorance entropy  of the reduced density matrix.  On the other hand the entanglement entropy crucially depends on reducing the density matrix -- it vanishes for a fixed configuration of the color charges $\rho_a(\v{q})$, but is nonzero for $\hat\rho_r$.
This is a clear indication that the ignorance entropy in general is not related with  entanglement.

\section{Conclusions}

In this work,  we have compared the entanglement entropy $S_E$ with the entropy of ignorance $S_I$ in a computable model. The entropy of ignorance, $S_I$  was defined as entropy associated with the fact that only limited number of observables is available for measurement in a quantum system. The model we have chosen has a number of similarities with the parton model of QCD.

Our comparison shows that in general $S_E$ and $S_I$ can be  quite different. In the context of the parton model $S_I$ is equal to the Boltzmann entropy of a classical ensemble of noninteracting partons. We found for example, that for a fixed configuration of the valence charges (analogous to fixed configuration of low transverse momentum modes in the hadron wave function) $S_E$ vanishes, while $S_I$ does not. Moreover for a typical configuration $S_I$ is very similar to its value for ensemble average.

There is however one striking feature of our result that needs to be understood. We found that with the reduced density matrix $\hat\rho_r$, for both Renyi and von Neumann the differences between $S_I$ and $S_E$ disappear in the ultraviolet cf. Eqs.~(\ref{leadingR},\ref{subleadingR},\ref{uvi},\ref{uve}). To get some insight into this let us first ask which states contribute the most to the entropy in the ultraviolet.

First we note that the eigenvalues $\rho_i$ of $\hat\rho_r$ at fixed small momentum $q^2\ll g^2\mu^2$  have hierarchical structure, so that  $\rho_0=1-\delta$, $\delta\ll 1$, while $\rho_{n\ge 1}\ll 1$, and $\rho_1\gg \rho_2\gg \rho_3\gg....$. Also, since $\hat\rho_r$ is normalized, we have $\delta=\sum_{i=1}^\infty \rho_i\approx \rho_1$. Thus only $\rho_0$ and $\rho_1$ contribute to entropy to leading order at small $q^2$.

Consider the Renyi entropy first. Since at large transverse momentum $|\v{q}|$, $R\sim 1/q^2$, it is obvious from Eqs.~(\ref{matel},\ref{matels}) that the largest matrix element of $\hat\rho_r$ is the one with $n=\beta=m=\alpha=0$, as we alluded to in Sect.~\ref{Sect:vN}.  The Renyi entropy of $\hat\rho_r$ is dominated completely by the contribution of this matrix element. Since this element is on the diagonal of $\hat\rho_r$, it of course also contributes the same amount to the Renyi entropy of ignorance. This is the reason why the UV leading behavior of $S_R$ and $S_R^I$ is the same. 

Note that this leading matrix element is the matrix element in the vacuum state at a given value of momentum. The equality of the leading contributions to $S_R$ and $S_R^I$ in the UV is thus a rather trivial effect, inasmuch as it does not actually probe the distribution of partons in the density matrix, but only the probability that no partons are present. Asking about parton distribution is asking about subleading corrections to entropy. 

 It is indeed easy to see that on the level of the first $1/q^2$ correction  $S_R$ and $S_R^I$ behave differently. The $1/q^2$ corrections to $S_R$ in Eq.~(\ref{matels}) originate from two types of matrix elements. First, there are diagonal contributions with $n=\alpha=1$ or $m=\beta=1$ , and the rest of $n,\ m,\ \alpha,\ \beta$ vanishing.  These terms contribute to $S_R$ and $S_R^I$ equally. Then there are non diagonal contributions to $S_R$ , which are banished from $S_R^I$: these are contributions non diagonal in the total particle number, e.g. $n=m=1$, $\alpha=\beta=0$ or $\alpha=\beta=1$, $n=m=0$.
  As it turns out the contributions of terms diagonal and non diagonal in the particle number are equal. Thus the first corrections to the leading term reflect the non diagonal nature of $\hat\rho_r$ versus diagonal $\hat\rho_I$ and are  different for $S_R$ and $S_R^I$.

 Now let us consider the von Neumann entropy. Here the situation is somewhat different. The largest eigenvalue of $\hat\rho_r$ does not necessarily give the largest contribution to $S_E$. For a hierarchical density matrix like our $\hat\rho_r$,  the von Neumann entropy is 
\begin{equation}
S_E=-\rho_0\ln\rho_0-\sum_{i=1}^\infty\rho_i\ln\rho_i\approx 
\delta  - \delta \ln \delta = 
\delta \ln \frac{e}{\delta} \, , 
\label{Eq:Se_expl}
\end{equation}
where the leading logarithmic contribution $\delta \ln \delta$  originates from $i=1$ in Eq.~\eqref{Eq:Se_expl} while the linear correction in $\delta$ is from the ``vacuum'' matrix element $i=0$.    
The eigenvalue $\rho_0$ corresponds roughly speaking to partonic vacuum state, while $\rho_1$ correspond to a single parton with longitudinal polarization, with $\rho_1=\frac{g^2\mu^2}{q^2}$ (this correspondence is only approximate, since as we know $\hat\rho_r$ is not actually diagonal in the particle number basis). Indeed Eq.~\eqref{Eq:Se_expl} (up to the overall factor that arises due to summation over colors and integration over the transverse plane) coincides with Eq.~(\ref{uve}).
  
  In this discussion $\rho_0$ and $\rho_1$ are the eigenvalues of $\hat\rho_r$. The difference between these eigenvalues and the first two diagonal matrix elements however is small. In particular, since $\rho_{02}\sim\delta^2$, we have $\rho_{00}=\rho_0+O(\delta^2);\ \ \ \ \rho_{11}=\rho_1+O(\delta^2)$
Therefore the contribution to the ignorance entropy due to these terms is
\begin{equation}
S_I(q^2)=S_E(q^2)+O(\delta^2\ln1/\delta)
\end{equation}
  which is indeed born out by Eqs.~(\ref{uvi},\ref{uve}).
  
  We conclude that the identical UV asymptotics of $S_I(q^2)$ and $S_E(q^2)$ is due to the small occupation numbers of partons at large $q^2$. Indeed, at intermediate and low momenta where the occupation numbers per unit phase space volume are of order unity the difference between the two types of entropies becomes significant, at the order of 50\%.
  We expect that the real parton model of QCD shares these features. At very large momenta the entanglement and ignorance lead to the same entropy, while at low $Q^2$ the resulting entropies should be different. This is likely to be unrelated to any nontrivial dynamics of the ``environment'' degrees of freedom, such as confinement but is just the consequence of low occupation number of partons at high momentum.

To summarize, our understanding is that  the lack of coherence and large entropy of the partonic density matrix within the parton model approach must be due to ``ignorance'', i.e. to our ability to measure only a restricted number of observables, rather than to the entanglement of the observed partons with the unobserved degrees of freedom, as suggested in Ref.~\cite{Kharzeev:2017qzs}.

\acknowledgements
We thank A. Kemper, E. Levin, and Z. Tu for illuminating discussions. 

We gratefully acknowledge support  by  NSF Nuclear Theory grant 1913890 (A.K.)  and by the US Department of Energy grant de-sc0020081 (H.D. and V.S.). 
V.S. thanks the ExtreMe Matter Institute EMMI (GSI
Helmholtzzentrum f\"ur Schwerionenforschung, Darmstadt, Germany) for partial support and hospitality.

\appendix 

\section{Entropies of entanglement and ignorance for a simple two fermion system}
As a simple example of calculation of the entropy of ignorance consider two fermions, A and B in the following pure state
\begin{align}
    |\phi_{AB}\rangle= \frac{\sqrt{2}}{2} |0_A\rangle \otimes |0_B\rangle +\frac{1}{2} |1_A\rangle \otimes \left(|0_B\rangle+ |1_B\rangle \right)\,.
\end{align}
Since this is a pure state, its  von Neumann entropy vanishes.

Let us calculate the standard entanglement entropy of a single particle subsystem.
After tracing out particles A or B, the reduced density matrix in the particle representation basis  for subsystem  A and B are
\begin{align}
	\rho_A= \frac{1}{2} \begin{pmatrix}
1 & \frac{\sqrt{2}}{2} \\
\frac{\sqrt{2}}{2}  & 1
\end{pmatrix}\,,\\
      \rho_B= \frac14 \begin{pmatrix}
3 & 1 \\
1 & 1
\end{pmatrix}\,.
\end{align}
The  entanglement entropies for the subsystem A and its complement are identical (as they should be)
\begin{align}
    S_E (\rho_A)= S_E (\rho_B) =  %=-(\frac{2+\sqrt{2}}{4}\ln(\frac{2+\sqrt{2}}{4})+\frac{2-\sqrt{2}}{4}\ln(\frac{2-\sqrt{2}}{4}))
    \frac{3}{2} \ln 2 + 
    \frac{1}{\sqrt{2}}\, {\rm acoth} \sqrt{2} \approx 0.416496\,.
\end{align}

The ignorance entropy depends on the set of defining operators $\{O_i\}$. Let us first take $\{O_i\}$ as all operators diagonal in the particle number basis. To calculate $S_I$ in this case we should take the density matrix discarding the off-diagonal matrix elements  in the number basis, $\rho_{AB}={\rm diag}\left\{{1}/{2}, \, {1}/{4},\, 0,\, {1}/{4}\right\}$ 
and 
\begin{align}
    S_I(\rho_{AB})=-\sum_ip_i\ln p_i=\frac{3}{2}\ln 2  \approx 1.03972\,.
\end{align}

Another simple quantity is the entropy of ignorance for the reduced density matrix $\rho_A$. This time the measurable quantities are operators diagonal in Fock space of fermion A. The diagonal density matrix is obtained by dropping the off-diagonal matrix elements of $\rho_A$: $\rho^I_A={\rm diag}\left\{1/2,\, 1/2\right\}$.
\begin{align}
     S_I(\rho_A)=\ln 2 \approx  0.693147\,.
\end{align}
Similarly, $\rho^I_B={\rm diag}\left\{3/4,\, 1/4\right\}$, and the corresponding entropy of ignorance is
\begin{align}
    S_I(\rho_B)=%-(\frac{3}{4}\ln(\frac{3}{4})+\frac{1}{4}\ln(\frac{1}{4}))
    2\ln 2 -\frac{3}{4}\ln 3  \approx  0.56233\,. 
\end{align}
Note that as opposed to the corresponding entanglement entropies, the two entropies of ignorance are not equal to each other  $S_I(\rho_A)\ne S_I(\rho_B)$.

\section{Mode sum for Renyi entropy}
\label{App}

Here we present the explicit form for the mode sum  $  \mathfrak{S}$:
\begin{align}
  \mathfrak{S}&=
   \sum_{m,n} 
  \left[\frac{\left( m+n  \right)!}{n!m!}
	  \left( \frac{R}{2} \right)^{m+n} 
 \right]^2\,.
 \label{Eq:sum_app}
\end{align}
Using the integral representation of $\Gamma$-function for $  \left[ (m+ n)! \right]^2   $
\begin{equation}
   \left[ (m+ n)! \right]^2 = \int_0^\infty dt_1 dt_2 e^{-t_1-t_2} (t_1 t_2)^{m + n} 
\end{equation}
alows us further to factorize the sums. After this factorization, we get 
\begin{align}
	\mathfrak{S}=\int_0^\infty dt_1 dt_2 e^{-t_1-t_2}  \left(  \sum_{m}  \frac{ 1 } { (m!)^2 } \left(\frac{R}{2}  \sqrt{t_1 t_2   } \right)^{2m} \right)^{2 }\,. 
\end{align}
Each of these sums gives  modified Bessel function $I_0$:  
\begin{align}
  \mathfrak{S}=\int_0^\infty dt_1 dt_2 e^{-t_1-t_2} 
    I_0^2\left( R \sqrt{t_1 t_2   }  \right) \,.
\end{align}
One of this integrals can be analytically computed after the change of variables $x=\sqrt{t_1 t_2}$ 
\begin{align}
    \mathfrak{S}&=2 \int_0^\infty dx x \ \int_0^\infty \frac{dt_1}{t_1} e^{-t_1-\frac{x^2}{t_1}} 
  I_0^2\left(R x \right) \notag \\
    &=4 \int_0^\infty  dx x K_0(2x)
     I_0^2\left(R x \right) \,.
\end{align}
The last equality is based on 10.32.10 from Ref.~\cite{NIST:DLMF}. 
Finally, the integral over $x$ can be done analytically; it is 0 for $|R|\ge 1$ and   
\begin{align}
    \mathfrak{S}&=
	\frac{1}{\sqrt{1-R^2} } 
\end{align}
otherwise.

\bibliography{EE}

\end{document}